# Reference vectors in economic choice

**Teycir Abdelghani GOUCHA**
University of Tunis, Tunisia
goucha@hotmail.com

**Abstract.** *In this paper the introduction of notion of reference vector paves the way for a combination of classical and social approaches in the framework of referential preferences given by matrix groups. It is shown that individual demand issue from rational decision does not depend on that reference.*

Introduction

A very important motive of economics has been the explanation of choices and decisions of agents in real life situations. To prove why there is equilibrium in free market where supply and demand are regulated by prices (Arrow, Debreu, 1954), or to explain failure of economic system during crisis and why it give rise to unemployment and recession (Taouil, 2004), one have recourse to the microeconomic approaches of consumers and producers decision.

To formulate logic consistent and practical model of free exchange market and to explain preferences and actual decisions of agents, utility functions and their derivatives have been mathematically useful and efficient during past decades. In such model, hypothesis of perfect rationality is assumed, in the sense that all agents tend to maximize their utility functions subject to budget constraint (Balasko, 1998). But despite its convenience and theoretical success, the concept of rationality was not able to integrate all complexities and diversities of agent's behavior. Some new ideas like information asymmetry, bounded rationality and uncertainty were developed in order to furnish a more realistic description of human actions involved in the economic world which is in perpetual expansion and transformation. At the same register, many approaches for exploring social and psychic dimensions in economic activity were elaborated (Simon, 1986, Keizer, 2007).

In this paper we establish a connection between classical and social approaches in the mathematical framework defined by referential preferences presented in previous work (Goucha, 2012). In the first paragraph we introduce the notion of reference vector which serves as a tool of comparison and assessment of commodities, and we analyze in the second how the choice of such vector may affect individual behavior. This analysis has allowed to build bridge between social and rational points of view of economic activity.

## 1. Referential preferences and vectors

A first definition of referential preferences (*RP*) was given by Goucha (2012). This new approach consists to classify bundles of commodities represented in $R^l_+$ by using matrix action's on the vector space. In fact, it is easy to check that any matrix subgroup $G \subseteq GL(l,R)$ induces an equivalence relation on $R^l$ defined as following:

$$\forall\, x,y \in R^l,\ x \sim y\ iff\ \exists\, M \in G\ st\ M \times x = y.$$

In microeconomics, equivalence relations determines indifference set of consumption, but since this is not sufficient to give a totally (complete) preorder on $R_+^l$ which is necessary to define preferences, some other conditions are needed.

**Axiom 1**. Let $G$ a matrix subgroup of $GL(l, R)$ with a globally invariant action on $R_+^l$.

For all $x \in R_{++}^l$, there is a unique $v_x \in R_+$ and $M_x \in G$ such that $x = v_x(M_x \times I)$, where $I = \begin{pmatrix} 1 \\ \vdots \\ 1 \end{pmatrix} \in R^l$.

We can deduce a preference relation on $R_{++}^l$ if axiom 1 is verified for some group $G$. Indeed, we say that $x$ is more desirable than $y$ when $v_x > v_y$, and they are equivalent if $v_x = v_y$.

We simply note that $v_x = v_y \Leftrightarrow \exists M \in G$ such that $M \times x = y \Leftrightarrow x \sim y$.

*Definition 1 (Referential preferences).* We say that a preference relation on $R_{++}^l$ is of reference type, or referential, whenever it is given by a matrix subgroup $G$ which satisfies axiom 1.

In many examples, axiom 1 is available for all $x \in R_+^l$ and the above definition can be extended to all commodities in $R_+^l$. If this is not the case we assume that all $x \in R_{++}^l$ are preferred to anything on the boundary.

The above axiom is not only a new mathematical formulation of a well-known notion, but it express an economic point of view which asserts that when an agent has to choose between two commodity bundles, he compares them to a third one which is familiar to him and more simple to estimate. This bundle represents a kind of personal reference value and is given by: $v \times I = \begin{pmatrix} v \\ v \\ \vdots \\ v \end{pmatrix}$.

Other information about preference$s$ given by axiom 1 reveals that the value $v_x$ attributed by agent to the bundle $x \in R_+^l$ depends on his proper vision of intrinsic relations between the $l$ commodities. For example, if he considers that $\begin{pmatrix} t & 0 \\ 0 & 1/t \end{pmatrix}\begin{pmatrix} 1 \\ 1 \end{pmatrix} \sim \begin{pmatrix} 1 \\ 1 \end{pmatrix}, \forall t > 0$, then he attributes **1** (1 as one unity of value) to the bundle $\begin{pmatrix} 2 \\ 1/2 \end{pmatrix}$. But if he believes that the matrix group $\left\{ \begin{pmatrix} t & 0 \\ 0 & 1/t^2 \end{pmatrix}, t > 0 \right\}$ is more adequate to describe relations in the consumption set, then he attributes to the same bundle the value $\sqrt[3]{2}$ since we have: $\begin{pmatrix} 2 \\ 1/2 \end{pmatrix} = \sqrt[3]{2} \begin{pmatrix} \sqrt[3]{4} & 0 \\ 0 & \frac{1}{\sqrt[3]{16}} \end{pmatrix} \begin{pmatrix} 1 \\ 1 \end{pmatrix}$.

This remark suggests that referential preferences lead to deduce values of bundles from the existing symmetries between commodities[1]. Among these symmetries naturally given by matrix subgroups of $GL(l, R)$, agent decides to choose one to formulate his preferences. At this record, Marin Dinu emphasizes in recent editorial the importance of internalization of symmetries intentionality in the configuration of economic reality (Dinu, 2013).

The role of symmetries being mentioned, we return with the following statement to the main theme of this paper which is obviously the reference vectors. We will show that if axiom 1 is valid for a group G, then it may be formulated with any other vector $X \in R_{++}^l$ instead of $I = \begin{pmatrix} 1 \\ \vdots \\ 1 \end{pmatrix}$.

**Proposition 1.** Let $G$ be a matrix subgroup which verifies axiom1 and a vector $X \in R_{++}^l$. Then for all $Y \in R_{++}^l$, there is a unique $v_y^x \in R_+$ and $M_y^x \in G$ such that $Y = v_y^x(M_y^x \times X)$.

**Proof:** By axiom1 there is $v_x, v_y \in R_+^*$, and $M_x, M_y \in G$ such that $X = v_x M_x \times I$ and $Y = v_y M_y \times I$. But since all matrices in G are invertible, then we have $I = \frac{1}{v_x} M_x^{-1} \cdot X$. This leads to the following equality:

$$Y = v_y \left( M_y \left( \frac{1}{v_x} M_x^{-1} \times X \right) \right) = \frac{v_y}{v_x} \left( M_y M_x^{-1} \times X \right),$$

or equivalently $Y = v_y^x M_y^x \times X$

where $v_y^x = \frac{v_y}{v_x}$ and $M_y^x = M_y M_x^{-1}$.

To prove the uniqueness, suppose that there is $u > 0$ and $M \in G$ with $(u, M) \neq (v_y^x, M_y^x)$ and $Y = u(M \times X)$. Consider the relation $X = v_x M_x \times I$, we obtain that $Y = u \times v_x (M \times M_x \times I)$. By axiom 1 and uniqueness of $v_y$ and $M_y$, we deduce that $u \times v_x = v_y$ and $M \cdot \times = M_y$ which states that $u = \frac{v_x}{v_y} = v_y^x$ and $M = M_y \times M_x^{-1} = M_y^x$.

The previous proposition suggests the new formulation of axiom 1:

*Axiom 1*. Let $G$ be a matrix subgroup with globally invariant action on $R_+^l$ and a vector $R \in R_{++}^l$. Then for all $x \in R_{++}^l$ there is a unique $v_x \in R_+^*$ and $M_x \in G$ such that $x = v_x(M_x \times R)$.

Referential preferences described by Axiom 1' are given in a same manner as axiom 1. We say that $x$ is more desirable than $y$ when $v_x > v_y$, and they are equivalent if $v_x = v_y$.

*Corollary.* Axioms 1 and $1'$ are equivalent.

*Proof:* According to proposition 1, axiom 1 implies axiom $1'$, and the converse follows by applying the latter to $I = \begin{pmatrix} 1 \\ \vdots \\ 1 \end{pmatrix}$.

In the light of what preceded, the following definition becomes not only possible but also well justified:

*Definition 2*. A vector $R \in R_{++}^l$ which verifies axiom 1' for some group $G$ is called a reference vector ($RV$), or more simply the referential.

At present, referential preferences are not only given by matrix subgroup $G$, but, unless to prove its neutrality in the decision process, it is also specified by a referential $R \in R_{++}^l$. The role of $R$ in the determination of individual demand and agent behavior will be the subject of the second paragraph.

## 2. References and choices in economic exchanges

For Simon, rationality in classical economics is viewed in terms of the choices it produces; while in the other social sciences it is viewed in terms of the processes it employs (Simon, 1986). What we make an attempt to give some answer here, is the relationship between the final choice and the process of decision in the case of individual demand. More precisely, how the reference vector which takes his place in the process can affect or not an agent's choice. To obtain a "first estimation" of response (term which come across further), let us compute and compare demand of three consumers having the same matrix subgroups of preferences and budget constraint, while their reference vectors are different.

*Example:* Consider a market with two commodities, $p = (\frac{1}{4}, \frac{3}{4})$ the price vector and $w=200$ the budget of three consumers with *RP* given by the matrix subgroup $G = \left\{ \begin{pmatrix} t & 0 \\ 0 & 1/t \end{pmatrix}, t \in R_+^* \right\}$ and by, $R_1 = \begin{pmatrix} 2 \\ 1 \end{pmatrix}$, $R_2 = \begin{pmatrix} 1 \\ 3 \end{pmatrix}$, $R_3 = \begin{pmatrix} 1 \\ 1/2 \end{pmatrix}$, as respective *RV*.

Since the preorder on the consumption set is given by $v = v_x$ where $x = v_x \times (M_x \times R)$, then the following problem determines consumer demand:

$$\begin{cases} \text{Maximize } v_x \\ \text{subject to the constraint } \langle p|x \rangle = w \end{cases}$$

*Consumer 1.* For all consumption bundle $x \in R_{++}^l$ there is a unique $v_x > 0$ and $M_x \in G$, such that $x = v_x \times (M_x \times R_1)$. Then to determine demand we have to resolve the following problem: Max $v_x$ such that $\langle \begin{pmatrix} 1/4 \\ 3/4 \end{pmatrix}, v_x \begin{pmatrix} t & 0 \\ 0 & \frac{1}{t} \end{pmatrix} \begin{pmatrix} 2 \\ 1 \end{pmatrix} \rangle = 200$ or equivalently, find Max $v_x$ such that $v_x \left( \frac{t}{2} + \frac{3}{4t} \right) = 200$.

A simple calculus gives that at $t = \sqrt{3/2}$ the maximal value of $v_x$ is reached and it is equal $v_{max} = \frac{200}{\sqrt{3/2}}$. With these data, we obtain that the demand of the first consumer is:

$$X_1 = \frac{200}{\sqrt{3/2}} \begin{pmatrix} \sqrt{\frac{3}{2}} & 0 \\ 0 & \frac{1}{\sqrt{\frac{3}{2}}} \end{pmatrix} \begin{pmatrix} 2 \\ 1 \end{pmatrix} = \begin{pmatrix} 400 \\ 400/3 \end{pmatrix}$$

*Consumer 2.* In a similar manner, we have to resolve the following problem of maximization: Max $v_x$ such that $\langle \begin{pmatrix} 1/4 \\ 3/4 \end{pmatrix}, v_x \begin{pmatrix} t & 0 \\ 0 & \frac{1}{t} \end{pmatrix} \begin{pmatrix} 1 \\ 3 \end{pmatrix} \rangle = 200$ or equivalently: Max $v_x$ such that $v_x \left( \frac{t}{4} + \frac{9}{4t} \right) = 200$. We obtain that $Max\ v_x = \frac{400}{3}$ given at $t = 3$ and the demand of consumer 2 is equal to:

$$X_2 = \frac{400}{3}\begin{pmatrix} 3 & 0 \\ 0 & \frac{1}{3} \end{pmatrix}\begin{pmatrix} 1 \\ 3 \end{pmatrix} = \begin{pmatrix} 400 \\ 400/3 \end{pmatrix}$$

*Consumer 3.* We have the following problem of maximization: Max $v_x$ such that $\langle \begin{pmatrix} 1/4 \\ 3/4 \end{pmatrix}, v_x \begin{pmatrix} t & 0 \\ 0 & \frac{1}{t} \end{pmatrix} \begin{pmatrix} 1 \\ 1/2 \end{pmatrix} \rangle = 200$ or equivalently: Max $v_x$ such that $v_x \left(\frac{t}{4} + 38t=200\right.$. For $t = 32$, the maximal value of $v$ is reached and we have Max $v_x = 400\sqrt{\frac{2}{3}}$ and the demand of consumer three follows as:

$$X_3 = 400\sqrt{\frac{2}{3}}\begin{pmatrix} \sqrt{\frac{3}{2}} & 0 \\ 0 & \frac{1}{\sqrt{\frac{3}{2}}} \end{pmatrix}\begin{pmatrix} 1 \\ 1/2 \end{pmatrix} = \begin{pmatrix} 400 \\ 400/3 \end{pmatrix}$$

In these three cases, individual demand remains the same what seems to suggest that referential has no impact on optimal choice of agent. We observe also that optimal value of $v_x$ took three different values. Economically speaking, agent satisfaction given by $Max\ v_x$ depends inter alia on its reference vector, but it is not, and more precisely, it should not be the case of its demand. However, since this observation is only based on examples, we first need a more general result proving the neutrality of *RV* in the optimization of individual choice. Actually, according to Goucha (2012), existence of demand function in the referential preference setting given by matrix subgroup *G* and $I = \begin{pmatrix} 1 \\ \vdots \\ 1 \end{pmatrix}$ as *RV* is subject to the following condition:

*Axiom 2.* There is a unique matrix $\bar{M} \in G$ such that: $0 < \langle I | \bar{M} \times I \rangle < \langle I | M \times I \rangle, \forall M \in G$.

It follows that demand function take this form: $f(p) = \frac{\langle p|e \rangle}{\langle I|\bar{M}\cdot I \rangle} P^{-1} \times \bar{M} \times I$, where $p = \begin{pmatrix} p_1 \\ \vdots \\ p_l \end{pmatrix}$ is the price vector, $e = \begin{pmatrix} e_1 \\ \vdots \\ e_l \end{pmatrix}$ the initial endowment of agent and *P* is a diagonal matrix with entries $P_{i,j} = \delta_{ij}\, p_j$ and $P^{-1}$ its inverse.

*Proposition 2.* The demand function of an agent with referential preferences giving by a matrix subgroup $G \subseteq GL(l, R)$ which verifies axiom 2 does not depend on the choice of reference vector $R \in R_{++}^l$.

*Proof:* At level price $p \in R_{++}^l$, and initial endowment $e \in R_+^l$, agent has to maximize $v_x$ such that $\langle p|x \rangle = \langle p|e \rangle$ where $x = v_x (M_x \times R)$. Since all terms are positive, and since axiom 1 implies that there is a unique $v_p > 0, v_r > 0$ and matrices $M_p, M_r$ in $G$ such that $P = v_p (M_p \times I)$ and $R = v_r (M_r \times I)$, we have then to minimize $v_p \times v_r \times \langle I | M_p \times M_x \times M_r \times I \rangle$ for $M_x \in G$. In virtue of axiom 2, this minimum which is clearly equal to $v_p \times v_r \times \langle I | \bar{M} \times I \rangle$, is reached when $M_p \times M_x \times M_r = \bar{M}$ or more precisely, when $M_x = M_p^{-1} \times \bar{M} \times M_r^{-1}$. From this we deduce that $\text{Max } v_x = \frac{\langle p|e \rangle}{v_p \times v_r \times \langle I|\bar{M}\cdot I \rangle}$ and the demand vector is $X = f(p) = \frac{\langle p|e \rangle}{v_p \times v_r \times \langle I|\bar{M}\cdot I \rangle} M_p^{-1} \times \bar{M} \times M_r^{-1} \times R$. By simply noticing that $I = \frac{1}{v_r}(M_r^{-1} \times R)$ and $\frac{1}{v_p} M_p^{-1} = P^{-1}$, we recover the expression $f(p) = \frac{\langle p|e \rangle}{\langle I|\bar{M}\cdot I \rangle} P^{-1} \times \bar{M} \times I$ which is independent from the choice of reference vector $R$.

Despite that proposition 2 incites to consider reference vector as neutral entity in the choice process, results of example suggest to examine the details of proof. In fact, we have seen that for a given price $p$ and a reference vectors $R \in R_{++}^l$, agent's demand is given by the bundle $X = \frac{\langle p|e \rangle}{v_p \times v_r \times \langle I|\bar{M}\cdot I \rangle} M_p^{-1} \times \bar{M} \times M_r^{-1} \times R$. Maximal reached value is, by consequence, a decreasing function of $v_r$. Stated otherwise, the less the reference is evaluated, the more the satisfaction or the feeling of satisfaction will be higher.

In the daily business practice, most buyers, especially professionals and intermediaries, tend to convince and persuade the seller (often without business experience) of the poor quality of its own merchandise. In doing so, they increase "margin of appreciation" while they seek to buy at the lowest price. We recover here the analysis of economic psychology where framing, which is here the final price to be reached, results from a modelled and well-intentioned impression (see Keizer, 2010, pp. 34-35). This influence is often exerted by referring to other goods which are both readily identifiable and having some proximity and resemblance to goods and services to be acquired or to sell. It is a kind of legitimation of the suggested price that one seeks rather to impose by the recourse to reference frames which have a certain recognized value. In all markets we often hear phrases such that: Sir, the apartment of your neighbour is on sale since three years... or, your car, Madam, is older than the YZ sold for less than $1000, etc.

That practice is also very common in the labour market where pressure on wages is justified by minimizing the contributions of employees and the value of their labour intake by some typical arguments in reference to other situations: we can earn more if we relocate our company or more explicitly: our direct competitors have cut year-end bonuses by half.

Accepting by someone the reference proposed by the other is equivalent to take as given a certain interval in which the amount of transaction will be located. Inspired by the social approach (Keizer, 2010, pp. 34-35), we can say that the framing is coming from impression or first estimate which one agent attempts to impose to other one. Whether the impression comes from some real data of market, or is just a sales tactic to increase gain, it does not fit into the topic of this paper. One of the main ideas expressed in this paper is the identification of referential as source of impression which we have also called the first estimate. The framing resulting from reference vector $R$ depends of the value $v_r \in R_+^*$ where $R = v_r M_r \times I$.

To conclude this brief discussion it should be noted that we just made some simple observations about the nature of $RV$ and its potential role in the choice process and effective decision made by economic actors. A more elaborate work should be conducted to improve our understanding of preferences and reference vectors and their impact on market exchange mechanisms. In agreement with proposition 2 and its interpretation, the starting point would be to look at rationality as an extreme hypothesis nuanced in practices.

## Conclusions

In this paper we establish a connection between the assumption of perfect rationality of economic agents, and other approaches that postulate psychic and social aspects. This connection established within the referential preferences framework allowed the identification of reference vectors as a common base of substantive and procedural analysis. Preliminary observations made in this article should be more clarified with other mathematical propositions and a deeper understanding of their economic and social interpretations.

## Note

[1]   This remark remains true if we consider relative value instead of the value itself.